# Method for Estimation of Nonlinear λ-I Relation of Single Phase Power Transformer Core by Using On-line Measurable Parameters

Farhad Gh. Khodaei, Iran
e-mails: st_f.khodaei@urmia.ac.ir, fkg1082@gmail.com

**Abstract**  In design of new power transformers, reliable and efficient tools are required to expedite research and development processes. Some of these tools are used to interpret the data obtained from the transformer tests for better judgement about the response of the transformer. In this paper a clear procedure is proposed to estimate the nonlinear relation between flux and excitation current of power transformers by using online measurable data, which is the main reason of the input current harmonics. It is shown that the procedure computes this relation with very high precision and leads to identical results to that applied in simulation.

**Keywords**  flux-current relation, harmonics, model estimation, non-linearities, power transformer.

## Introduction

Power transformer technology plays vital role in any power system. In research and development on new transformers, reliable and efficient softwares are required to reduce trial and error as much as possible. Some of these tools are used in making theory and practice results near to each other, for better judgement about what should be done in the future of research programs. Therefore, it is important to interpret the online measured data as well as possible [1]. There are several works done so far to interpret on-line obtained data of power transformers to extract important information such as the winding parameters [2-8], linear equivalent circuits [9-11] by methods such as least-square [7-10] or genetic [11,12] algorithms. Some of previous works dealing with practical estimation of core nonlinearities of transformers [13,14] do not use on-line parameters for estimation. Although [15,16] propose technique to model saturation response of transformer core at real-time (on-line) conditions, but the used techniques in these papers are based on some pre-assumptions about the transformer core nonlinear response. Also [17,18] propose estimation methods for hysteresis and don't focus on saturation response of transformers. In this paper, the main purpose is to extract a curve showing the nonlinear relation between flux and excitation current of transformers (or core saturation function), by using the parameter values obtainable from on-line tests, to study the quality of the iron cores used in fabrication of a power transformer in research and development process. This will be done by a new method compared to previous works, mainly based on nonparametric regression techniques.

## Description of the technique

The model used to estimate parameters of the transformer, is shown in Fig. 1. In this model the main part that is intended to be estimated in this paper, is the nonlinear inductor with voltage $v_x$ and current $i_m$. Other parameters such as the series impedances, should be determined by other tests prior to this parameter estimation procedure.

By writing two KVL equations, we have

$$\begin{cases} v_x(t) = v_2(t) + R_2 i_2(t) + L_2 \dfrac{di_2(t)}{dt} \\ v_1(t) = a v_x(t) + R_1 i_1(t) + L_1 \dfrac{di_1(t)}{dt} \end{cases} \quad (1)$$

That results in the following equations after computing the integrals of equations (1):

$$\begin{cases} \int_{-\infty}^{t} v_x(\tau)d\tau = \int_{-\infty}^{t} v_2(\tau)d\tau + R_2 \int_{-\infty}^{t} i_2(\tau)d\tau + L_2 i_2(t) \\ \int_{-\infty}^{t} v_1(\tau)d\tau = \int_{-\infty}^{t} v_{xp}(\tau)d\tau + R_1 \int_{-\infty}^{t} i_1(\tau)d\tau + L_1 i_1(t) \end{cases} \quad (2)$$

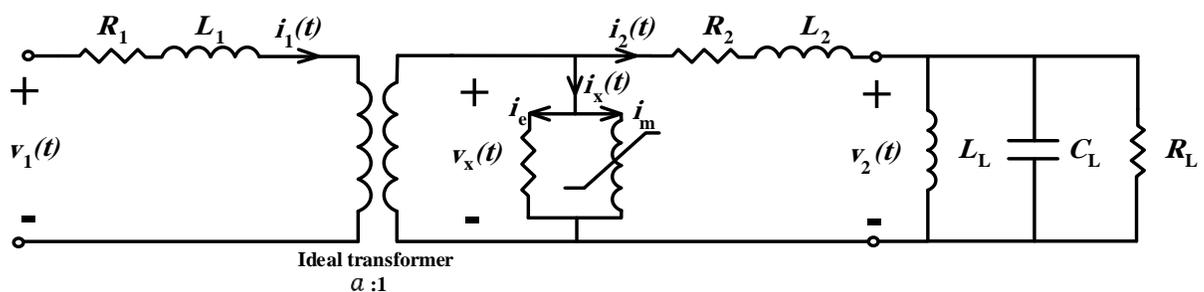

Fig. 1. A diagram for a real transformer model and definition of the used signals.

in which $v_{xp}(t)$ is defined as the primary voltage of the ideal transformer used in the model of Fig. 1 and satisfies the following equation:
$$v_{xp}(t) = a v_x(t) \quad (3)$$
Therefore, $i_x(t)$ can be computed as follows:
$$i_x(t) = a i_1(t) - i_2(t) \quad (4)$$
In order to do these computations with discrete signals (e. g. obtained by A/D converters), we define the integral of signal $f(t)$ in general, with the following equation:
$$F[n] = \int_{-\infty}^{t} f(\tau)d\tau \Big|_{t=nT_b} \cong T_b \sum_{k=-\infty}^{n} f(kT_b) \quad (5)$$
in which $T_b$ is the time between two adjacent samples of the signal $f(t)$. The following equation for $\gamma = 0$ can be used to compute the integral signal $F[n]$ by real-time samples of $f(t)$, but in order to prevent instability in the integral signal $F[n]$, a small and positive value for $\gamma$ can be chosen (e. g. 10$^{-4}$):
$$F[n] - (1-\gamma)F[n-1] = T_b f[n]$$
$$\text{with: } 0 < \gamma \ll 1 \quad f[n] \cong f(nT_b) \quad (6)$$
Now, if the integral signals of voltages and currents are defined by using upper-case letters, eq. (2) can be written as
$$\begin{cases} V_x[n] = V_2[n] + R_2 I_2[n] + L_2 i_2[n] \\ V_{xp}[n] = V_1[n] - R_1 I_1[n] - L_1 i_1[n] \end{cases} \quad (7)$$
If voltage and current of the terminals of the transformer are measured, the signals at the right sides of (7) will be known. Therefore, the left side signals can be computed. This means that the voltage signals of $v_x[n]$ and $v_{xp}[n]$ can be computed by using the inverse of Integral process as follows:
$$v_x[n] = \frac{V_x[n] - V_x[n-1]}{T_b}, v_{xp}[n] = \frac{V_{xp}[n] - V_{xp}[n-1]}{T_b} \quad (8)$$
Here, because at low absolute values of $v_x[n]$ and $v_{xp}[n]$, these signals may be slightly noisy, the conversion ratio of the ideal transformer of the model or $a$, can be computed by using maximum of absolute values of $v_x[n]$ and $v_{xp}[n]$ as follows:
$$a = \max(|v_{xp}[n]|)/\max(|v_x[n]|) \quad (9)$$
By using a known value for $a$, the current signal $i_x[n]$ can be computed as follows:
$$i_x[n] = a i_1[n] - i_2[n] \quad (10)$$
Since at no-load conditions, the power factor seen from the primary is very low, and the currents $i_e$ and $i_m$ does not change significantly at full-load conditions compared to no-load, we can use the following approximation about currents $i_e$ and $i_m$:
$$i_x = i_e + i_m \cong i_m \quad (11)$$
However, if in somewhere, it is required to do computation with a high precision, the resistor $R_c$ that models the core losses (both eddy current and hysteresis) can be computed as follows:
$$\begin{cases} R_c = \frac{V_{rms}^2}{P} = E(v_x^2)/E(i_x v_x) \\ i_m = i_x - v_x/R_c \end{cases} \quad (12)$$
in which $E(.)$ represents mean value. This helps to compute a more exact value for the signal $i_m[n]$. Now, our main purpose can be described which is modelling of the nonlinear inductor in the model of the transformer, that maintains a relation between $V_x[n]$ and $i_m[n]$ (or $\lambda$ and I) by a function named $h$ as follows:
$$\int_{-\infty}^{t} v_x(\tau)d\tau = \lambda(t) = h(i_m) \quad (13)$$
or equivalently
$$V_x[n] = h(i_m[n]) \quad (14)$$
By a known definition for the function $h$, one of the signals $V_x[n]$ and $i_m[n]$ can be computed by the other, but the main purpose is to compute $h$ by using known $V_x[n]$ and $i_m[n]$ signals. This problem in general, is a Nonparametric Regression problem [19], but more simple techniques can be used to solve it, defined as follows:
$$\text{for cycle } k: p \in \mathbb{Z}, \frac{p}{100} \leq \frac{i_m[n]}{\sqrt{2}I_{b2}} < \frac{p+1}{100}$$
$$\Rightarrow h_k^o\left(\frac{p}{100}\right) = \frac{\omega V_x[n]}{\sqrt{2}V_{b2}} \quad (15)$$
in which $\omega$ is the power line angular frequency that can be $100\pi$ or $120\pi$ and $V_{b2}, I_{b2}$ are the base RMS values of the voltage and the current at the secondary side, in per-unit system. Also the argument parameter $x$ in $h_k^o(x)$ is defined as the instantaneous value of $i_m$ in per-unit system and $h_k^o(x)$ is defined as instantaneous value of $\lambda$, per unit, at cycle $k$ of the 50 or 60 Hz electrical power signal. Although $h_k^o(x)$ is obtained for each cycle individually, the integrals has effects of the signals from previous cycles because the integrals are computed from $-\infty$ to the present instant. According to (15), the function $h_k^o(x)$ is obtained for 1 percent discrete steps of $i_m$ in per-unit system. Then the obtained $h_k^o(x)$ at various cycles, can be combined by a low-pass system to prevent sudden changes in the resulted response as follows:
$$(1+\beta)h_k(x) - \beta h_{k-1}(x) = h_k^o(x) \quad \text{with: } \beta \gg 1 \quad (16)$$
and finally we obtain
$$h_{result}(x) = \lim_{k \to +\infty} h_k(x) \quad (17)$$

**Verification of the technique**

In order to verify the presented procedure, we use Simulink®, and then use MATLAB codes to process the resulted data. A schematic was used in Simulink to simulate a saturable power transformer shown in Fig. 2. A single-phase power transformer with a saturable core and voltage ratio of 22kV/400V was simulated and the voltage and current curves are shown in Figs. 3 and 4. The obtained current waveform of the nonlinear inductor in the model of the transformer shown in Fig. 4 is not sinusoidal and contains high level of harmonics.

In Fig. 5 the estimated data of $h_{result}(x)$ and the data used in simulation are shown and compared. It is seen in Fig. 5 that the estimated core saturation function or $h_{result}(x)$ is identical with the data inserted to the simulation as the core saturation response.

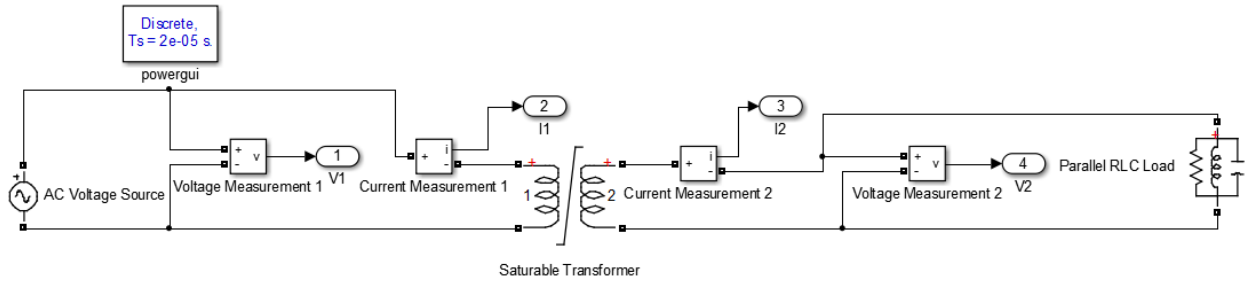

Fig. 2. A schematic in Simulink®, for simulation of a power transformer with a saturable and lossy core.

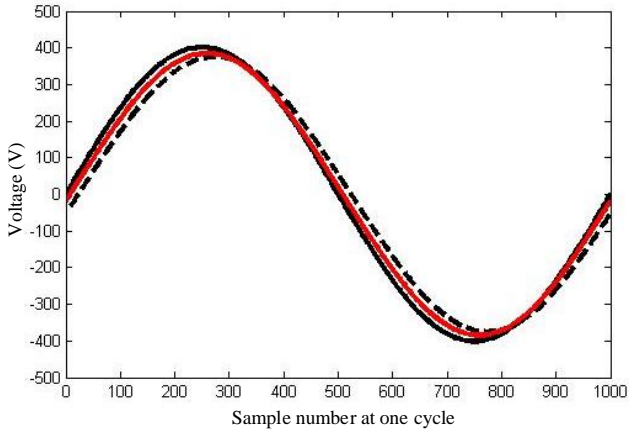

Fig. 3. Voltage waveforms of the primary side (applied by AC voltage source in Fig. 2) transferred to the secondary (solid black line), voltage of the secondary side of the simulated transformer (dashed line) and the obtained $v_x$ by the proposed technique (red line).

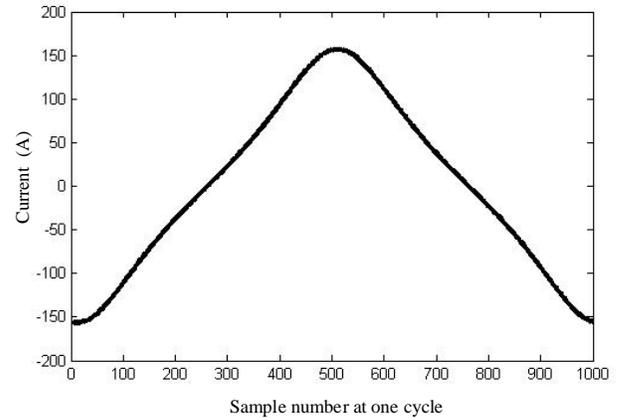

Fig. 4. The current waveform of the nonlinear inductor in the model of the transformer ($i_m$) obtained by the proposed technique.

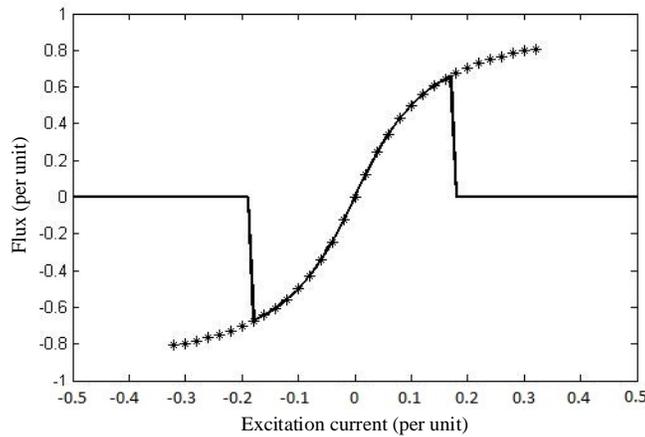

Fig. 5. The estimated core saturation function $h_{result}(x)$ (solid line) and the data inserted in Simulink as core $\lambda$-I relations (star points), note that there is not a phase difference between $\lambda$ and I because the core losses have been removed and considered by $R_c$ in eq. (12) and the focus of this paper is only the core nonlinearities (the solid line drops down at about 0.6pu flux strength because the initial guess of the core saturation function is zero).

## Conclusion

In this paper, a clear procedure was proposed to estimate nonlinearities of saturable power transformer core by using online obtainable voltage and current data. It was verified successfully by using Simulink and MATLAB tools. Using simulation helps to apply any wished model for the nonlinear inductor for better checking the accuracy of the technique, which is not possible in measurements. The main applications of this procedure can be in determination of maximum voltage ratings of a fabricated transformer, comparison of quality of iron cores of transformers in order to reduce current harmonics, or real-time monitoring of any changes in behavior of the transformer core, at real conditions such as transformer over-heating.

## Acknowledgement

The MATLAB and Simulink files of this project have been shared in GitHub address of
https://github.com/farhadkhodaei/transformer-model-estimation